\documentclass[12pt]{article}
\usepackage{graphicx}
\usepackage{amssymb}
\usepackage{amsmath}

\usepackage{setspace}
\usepackage{epstopdf}
\usepackage{bm}
\usepackage{float}

\def\bea{\begin{eqnarray}}
\def\eea{\end{eqnarray}}
\def\be{\begin{equation}}
\def\ee{\end{equation}}

\begin{document}

\title{Doublon dynamics and polar molecule production \\ in an optical lattice}
\author{Jacob P. Covey$^{1,2}$, Steven A. Moses$^{1,2}$, Martin G\"arttner$^{1,2}$, Arghavan Safavi-Naini$^{1,2}$, \\Matthew T. Miecnikowski$^{1,2}$, Zhengkun Fu$^{1,2}$, Johannes Schachenmayer$^{1,2}$, \\Paul S. Julienne$^{3}$, Ana Maria Rey$^{1,2}$, Deborah S. Jin$^{1,2}$, Jun Ye$^{1,2}$\\
\\
\normalsize{$^{1}$JILA, National Institute of Standards and Technology and University of Colorado,} \\ \normalsize{Boulder, CO 80309, USA} \\ \normalsize{$^{2}$Department of Physics, University of Colorado, Boulder, CO 80309, USA}
\\ \normalsize{$^{3}$Joint Quantum Institute, University of Maryland and National Institute of Standards and Technology,} \\ \normalsize{College Park, Maryland 20702, USA}}
\date{}

\maketitle

\begin{abstract}
Ultracold polar molecules provide an excellent platform to study quantum many-body spin dynamics, which has become accessible in the recently realized low entropy quantum gas of polar molecules in an optical lattice. To obtain a detailed understanding for the molecular formation process in the lattice, we prepare a density distribution where lattice sites are either empty or occupied by a doublon composed of a bosonic atom interacting with a fermionic atom. By letting this disordered, out-of-equilibrium system evolve from a well-defined initial condition, we observe clear effects on pairing that arise from inter-species interactions, a higher partial wave Feshbach resonance, and excited Bloch-band population. When only the lighter fermions are allowed to tunnel in the three-dimensional (3D) lattice, the system dynamics can be well described by theory. However, in a regime where both fermions and bosons can tunnel, we encounter correlated dynamics that is beyond the current capability of numerical simulations. Furthermore, we show that we can probe the microscopic distribution of the atomic gases in the lattice by measuring the inelastic loss of doublons. These techniques realize tools that are generically applicable to heteronuclear diatomic systems in optical lattices and can shed light on molecule production as well as dynamics of a Bose-Fermi mixture.
\end{abstract}

Polar molecules with long-ranged dipolar interactions are ideally suited to the exploration of strongly correlated quantum matter and intriguing phenomena such as quantum magnetism, exotic superfluidity, and topological phases~\cite{Baranov2008, Pupillo2009, Carr2009, Lahaye2009, Lemeshko2013,Yao2012,Gorshkov2011,Syzranov2014}. The recent observation of the dipole-mediated spin-exchange interaction in an optical lattice~\cite{Yan2013} and the demonstration of the many-body nature of the spin-exchange dynamics~\cite{Hazzard2014} mark important steps for the use of polar molecules to study strongly correlated matter. While this initial work was done with a molecular filling fraction of only $\sim$$5\%$ in a 3D lattice~\cite{Yan2013}, more recent work demonstrated a quantum synthesis scheme for molecule production in the lattice that relies on careful preparation of the initial atomic gases~\cite{Moses2015}. This led to a reduction in the final entropy of polar molecules by a factor of $\sim$4, and, correspondingly, a much higher filling fraction of $\sim$$25\%$ that opens up the possibility for studying non-equilibrium, many-body spin dynamics in a fully connected quantum gas of polar molecules.

The quantum synthesis approach reported in Ref.~\cite{Moses2015} starts by preparing atomic insulator states that depend on atomic interactions, quantum statistics, and low temperature~\cite{Safavi-Naini2015}.  However, realizing the full potential of this approach requires not only control over the atomic distributions, but also a detailed understanding of the molecule creation process.   Here, we investigate this important step by leveraging our capability of molecule production in an optical lattice to create a clean system of doublons~\cite{Winkler2006}. After creating ground-state molecules, we efficiently remove all unpaired atoms from the lattice and convert the molecules back to free atoms. This realizes a lattice where the sites are either empty or occupied by individual doublons that comprise a pair of bosonic and fermionic atoms.  This well-defined initial state allows us to directly address limitations in the molecule creation process by probing the efficiency with which these doublons are converted back to molecules under various experimental conditions that affect atomic tunneling rates, higher Bloch-band populations, and the adiabaticity of a magnetic-field sweep through a higher partial-wave Feshbach resonance. Additionally, we can use the well-initialized, non-equilibrium state of a disordered doublon distribution to explore the many-body dynamics of a lattice-confined Bose-Fermi mixture in a regime that is beyond the current simulation capabilities.

\begin{figure}
\centering
\includegraphics[width=7in]{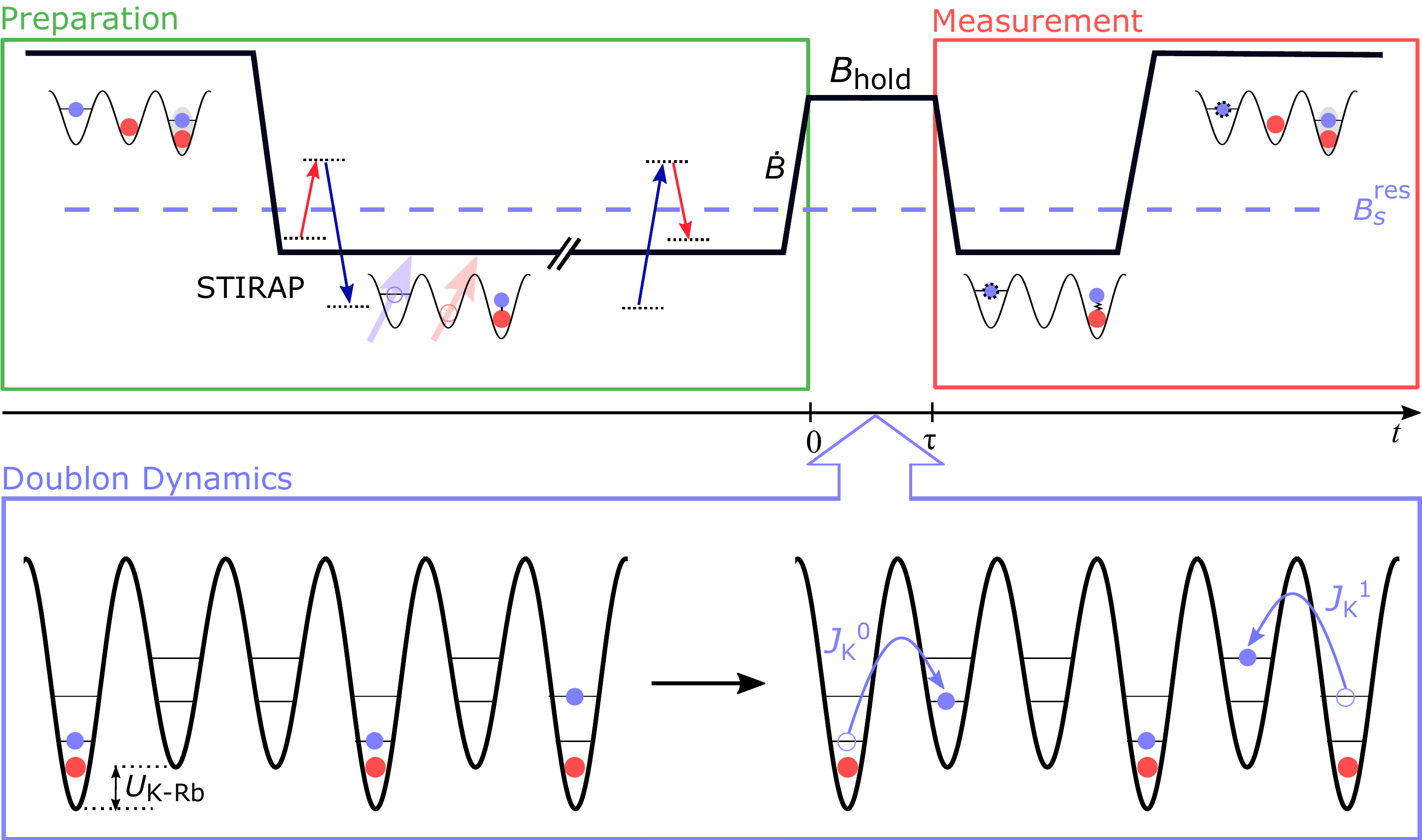}
\caption{\textbf{A schematic of the experiment}. Starting with a mixture of K, Rb, and doublons (the smaller blue ball, the larger red ball, and the pair grouped with grey background, respectively) in a 3D lattice, we sweep the magnetic field from above the $s$-wave Feshbach resonance (at $B_s^\text{res}$=54.66~mT) to below the resonance to create Feshbach molecules. These molecules are then transferred to their ro-vibrational ground state via STIRAP (stimulated Raman adiabatic passage). After unpaired atoms are removed with resonant light, the STIRAP process is reversed to transfer the ground-state molecules back to Feshbach molecules. The field is then swept above $B_s^\text{res}$ to dissociate the molecules and create doublons. After holding for a time, $\tau$, at $B_{\text{hold}}$,
we measure the conversion efficiency when sweeping the field below $B_s^\text{res}$ to re-form Feshbach molecules. To detect molecules, we use a rf pulse to spin flip the unpaired K atoms to a dark state (ball with black dashed edge)
before dissociating the Feshbach molecules and imaging K atoms. The bottom panel illustrates possible dynamics of the doublons during $B_{\text{hold}}$. As shown schematically, lattice sites populated with a K and a Rb atom have an interaction energy shift $U_{\text{K-Rb}}^0$. The K tunneling energies in the lowest and first excited bands are denoted by $J_\text{K}^0$ and $J_\text{K}^1$, respectively. Rb tunneling happens at a slower rate since it experiences a deeper lattice. 
}
\label{Figure1}
\end{figure}

The experiment proceeds in steps as depicted schematically in Fig.~\ref{Figure1}. To prepare the doublons, we create a sample of molecules in their ro-vibrational ground state in the lattice as described in Ref.~\cite{Moses2015} and then remove unpaired atoms with resonant light, so that all lattice sites are either empty or contain a single molecule. We then transfer the ground-state molecules back to a weakly bound Feshbach molecule state, followed with a magnetic-field ($B$) sweep to above the resonance to create a clean system of doublons. The solid black line in the upper panel of Fig.~\ref{Figure1} shows schematically $B$ relative to the $s$-wave Feshbach resonance (dashed line) that is used to manipulate the atomic inter-species interactions and to create molecules. After this preparation, the doublons are left to evolve in the lattice for a variable time $\tau$. Our measurement then consists of sweeping $B$ below resonance to associate atoms into Feshbach molecules and determining the fraction of K atoms that form molecules. Specifically, we measure the molecule number in the following protocol. We first apply rf to spin-flip the unpaired K atoms to another hyperfine state, which renders the unpaired K atoms invisible for subsequent molecular detection. We then sweep $B$ back above the resonance to dissociate the molecules, and measure the number of resulting K atoms with spin-selective resonant absorption imaging. The conversion efficiency is determined by dividing this molecule number by the total number of K atoms measured when we do not apply the rf.

\begin{figure}
\centering
\includegraphics[width=7in]{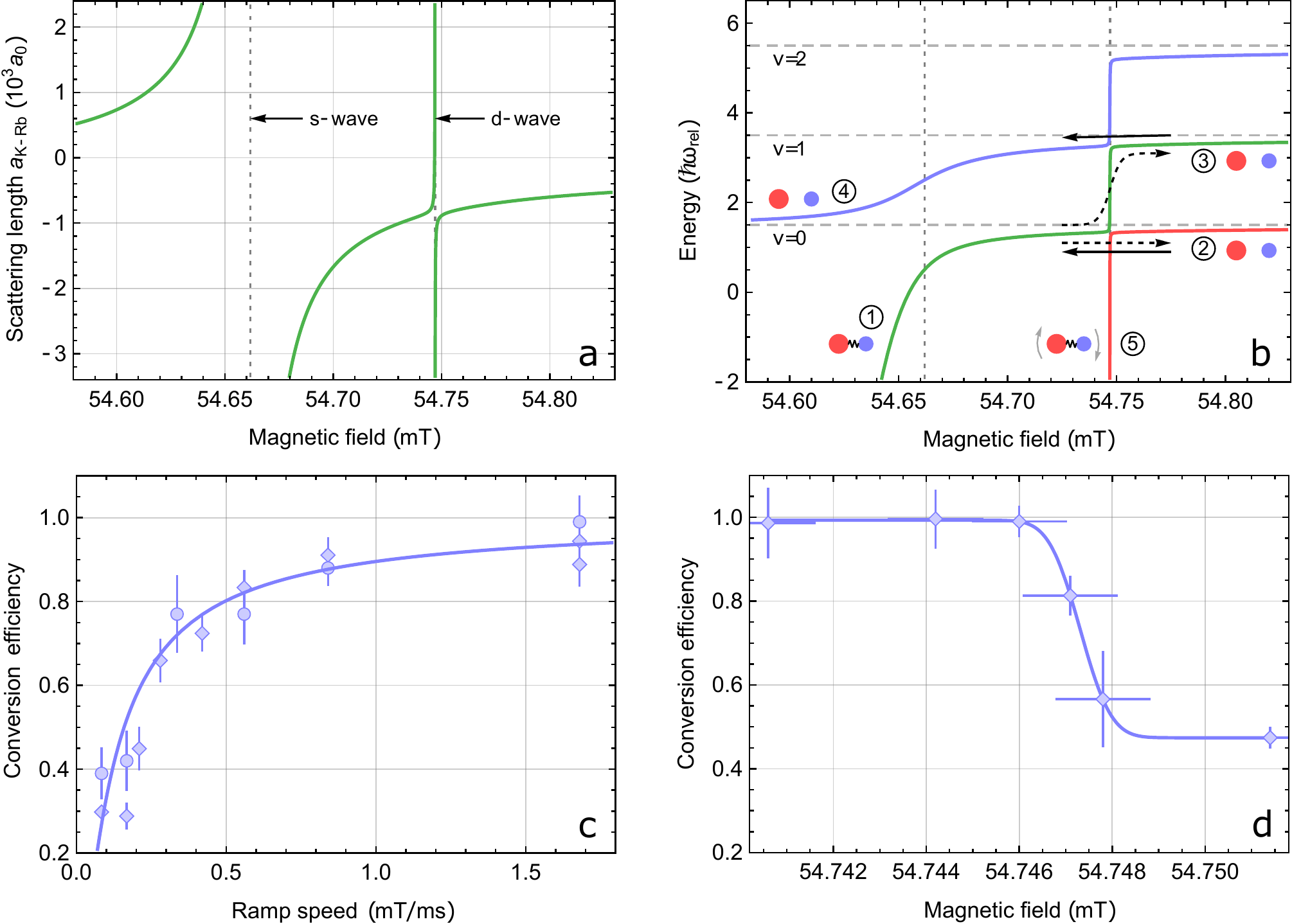}
\caption{\textbf{The $d$-wave Feshbach resonance}. \textbf{a}, The theoretical K-Rb scattering length, $a_{\text{K-Rb}}$, is shown as a function of the magnetic field for the broad $s$-wave Feshbach resonance and a narrow $d$-wave resonance, based on the formula and parameter values described in Methods. \textbf{b}, Crossing the $d$-wave resonance affects the pair states for K and Rb. Dashed and solid arrows shows the effect of the variable rate sweep that creates doublons and the subsequent fast magneto-association sweep, respectively. Dashed vertical lines mark the positions of the Feshbach resonances. 
 \textbf{c}, Measurement of molecule conversion efficiency at 35 $E_{\text{R}}$ (circles) and 30 $E_{\text{R}}$ (diamonds), with the latter data exponentiated by $(35/30)^{3/4}=1.12$ to account for the expected dependence on lattice depth. The solid curves shows a fit to a Landau-Zener avoided crossing, which gives a resonance width of $9.3(7)\times10^{-4}$~mT. \textbf{d}, The magnetic field at which this resonance occurs is determined by sweeping up  to various fields at 0.018 mT/ms, then sweeping down at 0.18 mT/ms. The position of the resonance extracted from this measurement at $V_\text{latt}=35E_\text{R}$ is 54.747(1) mT.}
\label{Figure2}
\end{figure}

We begin by investigating a narrow $d$-wave Feshbach resonance~\cite{Zaccanti2006, Julienne2009,  Bloom2013, Ruzic2013} that is located less than 0.1~mT above the 0.3~mT-wide $s$-wave resonance that is used for making molecules (Fig.~\ref{Figure2}a).  This narrow Feshbach resonance can adversely affect the magneto-association process, where $B$ is swept down from above the $s$-wave resonance to create molecules. Crossing the $d$-wave resonance too slowly will produce $d$-wave molecules, which will not be coupled to the ground state by the subsequent STIRAP laser pulses. If $B$ is swept sufficiently fast to be diabatic for this narrow resonance (but still slow enough to be adiabatic for the broad $s$-wave resonance), crossing the $d$-wave resonance has no impact; however, the high effective densities at each site in an optical lattice can make it challenging to sweep fast enough. Although we study here specific resonances for the K-Rb system, the possibility of having to cross other Feshbach resonances and the issue of sweep speeds are general to magneto-association of atoms in an optical lattice.

In the experiment illustrated in Fig.~\ref{Figure1}, we investigate the $d$-wave resonance by varying the rate, $\dot{B}$, and the final value, $B_{\text{hold}}$, of the sweep that creates doublons. We then measure the subsequent molecule conversion efficiency after $\tau=1$~ms using a fast 1.68~mT/ms magneto-association sweep. Figure~\ref{Figure2}b illustrates relevant states, above and below the resonance, for two atoms in a lattice site: at low fields, these are the $s$-wave molecule $\textcircled{1}$, $d$-wave molecule $\textcircled{5}$, and unbound atoms $\textcircled{4}$ and at high fields, unbound atoms in the ground band of the lattice $\textcircled{2}$ and atoms with a band excitation in their relative motion $\textcircled{3}$. For simplicity, we illustrate states for a harmonic potential whose trap frequency $\omega$ is the same for both atoms, with eigenstates of relative motion denoted by $v=0,1,2$. The dashed arrows show the diabatic ($\textcircled{1}\rightarrow\textcircled{2}$) and adiabatic ($\textcircled{1}\rightarrow\textcircled{3}$) trajectories for the dissociation of $s$-wave Feshbach molecules when crossing the $d$-wave resonance, while the solid arrows show the diabatic trajectories  ($\textcircled{2}\rightarrow\textcircled{1}$ and $\textcircled{3}\rightarrow\textcircled{4}$) for the subsequent, fast magneto-association sweep.

Figure~\ref{Figure2}c shows the measured molecule conversion efficiency as a function of  $\dot{B}$ when sweeping across the $d$-wave resonance to 56.24 mT, while  Fig.~\ref{Figure2}d shows the effect of the final $B$ field for a relatively slow, 0.018~mT/ms, sweep. The data are taken for lattice depths of $V_{\text{latt}} = 35 E_{\text{R}}$ (circles) and $30 E_{\text{R}}$ (diamonds), where $E_{\text{R}}=\hbar^2 k^2/(2 m_\text{Rb})$ is the recoil energy for Rb, $m_\text{Rb}$ is the Rb atom mass, $k=2\pi/\lambda$, and $\lambda=1064$~nm.
For our highest sweep rates, or when $B_{\text{hold}}$ is below the $d$-wave resonance, the measured molecule conversion efficiency is near unity.  This high conversion of doublons~\cite{Thalhammer2006,Greif2011,Chotia2012} is crucial for the quantum synthesis approach to producing molecules with a high filling fraction in the lattice. The near unity conversion also provides an excellent starting point for diagnosing potential limitations to molecule production, and the data in Fig.~\ref{Figure2}c and \ref{Figure2}d clearly show the negative effect that the $d$-wave resonance can have on magneto-association in the lattice.

The lines in Fig.~\ref{Figure2}c and \ref{Figure2}d show fits used to extract the width ($\Delta_d$) and position of the $d$-wave resonance. We use a Landau-Zener formalism~\cite{Mies2000} where the probability to cross the $d$-wave resonance diabatically, and therefore create $s$-wave Feshbach molecules in the subsequent magneto-association step, is $P = \exp \left[ -A/|\dot{B} |\right]$, where $A$ depends on the on-site densities and the Feshbach resonance parameters. By approximating the sites in the deep optical lattice as harmonic oscillator potentials, we extract $\Delta_d$ using $A = \frac{4 \sqrt{3} \omega_{\text{HO}}|a_{\text{bg}} \Delta_d|}{ L_{\text{HO}}}$~\cite{Julienne2004withnote}, where 
$\omega_{\text{HO}}$ is the harmonic trap frequency for relative motion of the two atoms (see Methods) and $L_{\text{HO}}=\sqrt{\hbar/(\mu \omega_{\text{HO}})}$ is the harmonic oscillator length with the doublon reduced mass $\mu$. From an exponential fit (line in Fig.~\ref{Figure2}c) value of $A=0.110(7)$~mT/ms, and using a background scattering length of $a_{\text{bg}}=-187(5)a_0$~\cite{Klempt2008}, where $a_0$ is the Bohr radius, we extract a width of $\Delta_d=9.3(7)\times10^{-4}$~mT. By fitting an error function (line) to the data in Fig.~\ref{Figure2}d, we determine the location of the resonance to be 54.747(1)~mT, which is consistent with previous experiments where atom loss was observed~\cite{Zaccanti2006, Bloom2013}.

The precise determination of the width of the $d$-wave resonance allows us to gauge its significance in molecule creation. Our typical sweep rate of 0.34~mT/ms for magneto-association, which has remained the same since the first creation of KRb molecules in an optical lattice~\cite{Chotia2012}, gives $\sim70\%$ probability of being diabatic when crossing the $d$-wave resonance. This suggests that we create a substantial fraction of $d$-wave molecules that are dark to our detection ($\textcircled{5}$ in Fig.~\ref{Figure2}b). These $d$-wave molecules may have played a role in limiting the lattice filling fraction for polar molecules achieved in Ref.~\cite{Moses2015}.

Tunneling dynamics of doublons in the lattice~\cite{Jurgensen2014} can also affect molecule production.  In the quantum synthesis approach, achieving a high lattice filling for molecules requires not only the preparation of a large fraction of lattice sites that have doublons but also that these doublons are not lost due to tunneling and/or collisions prior to conversion to molecules. In our system, K feels a lattice depth that, in units of recoil energy, is 2.6 times weaker than for Rb due to differences in atomic mass and polarizability. Consequently, K tunnels faster than Rb. While a sufficiently deep lattice can prevent tunneling of both K and Rb, practically this may not be possible in all cases, especially for polar molecule production using two atomic species that have large differences in mass and polarizability.

\begin{figure}
\centering
\includegraphics[width=7in]{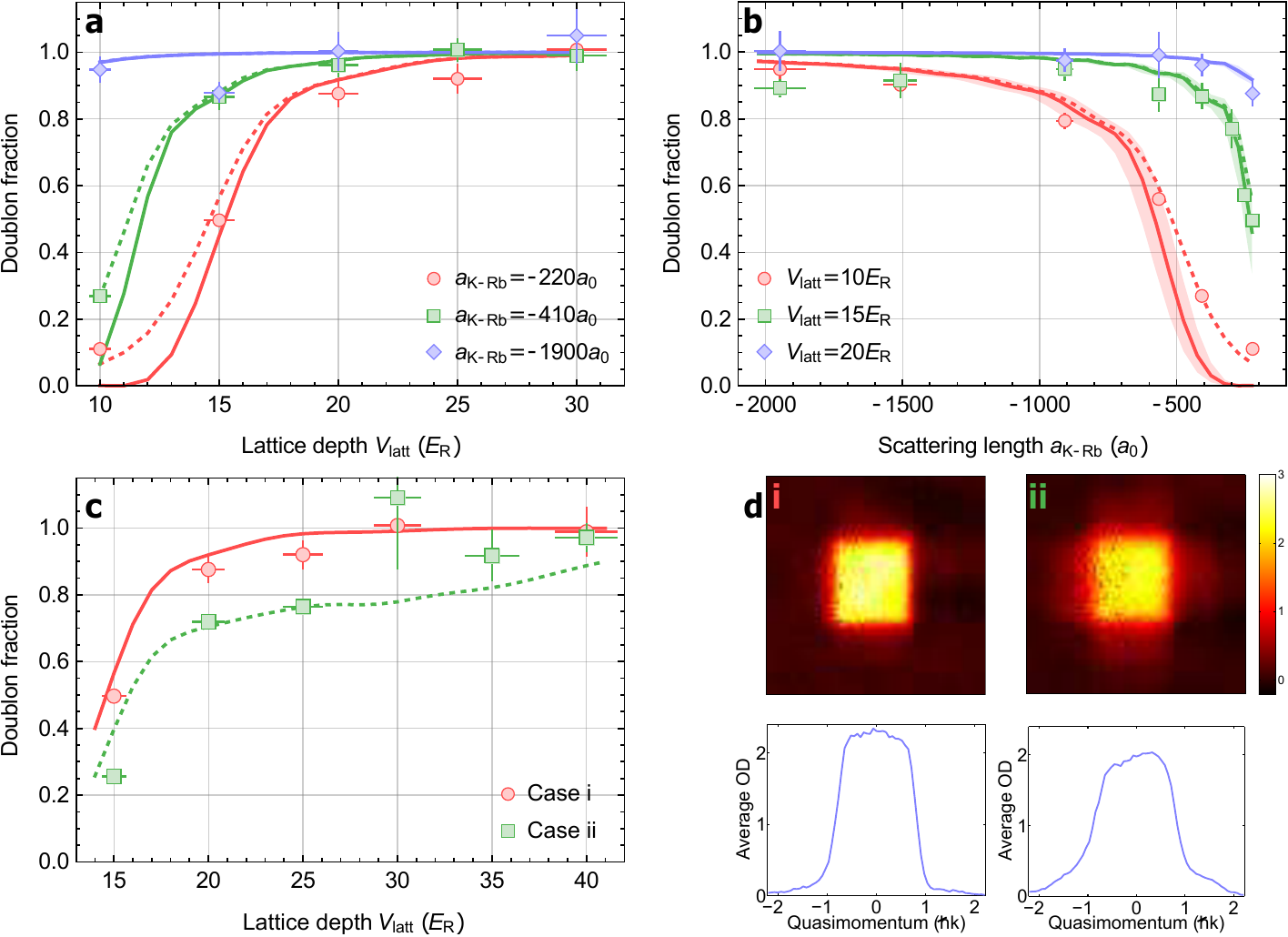}
\caption{\textbf{Interaction and tunneling dynamics of doublons in the lattice}. \textbf{a}, The remaining doublon fraction is shown for three scattering lengths as a function of lattice depth. \textbf{b}, The doublon fraction is plotted for three lattice depths as a function of the scattering length. \textbf{c}, The doublon fraction for 1.68~mT/ms sweeps, $\tau=1$~ms, and $a_\text{K-Rb}= -220a_0$ is shown as a function of lattice depth for the case of higher excited-band fraction (squares) and lower excited-band fraction (circles). \textbf{d}, Band-mapping images of the initial K gas are shown for the two different initial temperatures, where image \textbf{i} corresponds to the red circle data points and \textbf{ii} corresponds to the green square data points in \textbf{c}. Each image is the average of three measurements. Below the images, we show the optical depth for a horizontal trace through the image, with averaging from -$\hbar k$ to +$\hbar k$ in the vertical direction.
}
\label{Figure3}
\end{figure}

Figure~\ref{Figure3} illustrates doublon dynamics due to the interplay between tunneling and interactions, which we control by varying the lattice depth, interspecies scattering length $a_{\rm K-Rb}$, and band population. The fraction of doublons that remain after $\tau$ is essentially equal to the measured molecule conversion efficiency described above.  We note that for $a_\text{K-Rb}>-850 a_0$, the $B$ sweep crosses the $d$-wave Feshbach resonance with a $\dot{B}$ that varies from 0.5~mT/ms to 1.9~mT/ms. Using our measured width of the $d$-wave resonance, the data presented in Fig.~\ref{Figure3} have been multiplied by a factor that increases the doublon fraction to account for the finite $\dot{B}$ when crossing the $d$-wave resonance. Fig.~\ref{Figure3}a shows the effect of the lattice depth for $\tau=1$~ms at three different values of $B_{\text{hold}}$, corresponding to different $a_\text{K-Rb}$. We observe that the remaining doublon fraction is highly sensitive to the lattice depth for weak interspecies interactions, e.g. $a_\text{K-Rb}=-220a_0$, with a lower doublon
fraction for shallower lattices with higher tunneling rates. For stronger interactions, the dependence on lattice depth becomes less significant and almost disappears in the strongly interacting regime, e.g. $a_\text{K-Rb}=-1900a_0$. Similar behavior is observed if we fix the lattice depth but vary the interspecies interactions, as shown in Fig.~\ref{Figure3}b.

The data in Fig.~\ref{Figure3} clearly show evidence of decay of doublons due to tunneling that is affected by both the lattice depth and interspecies interactions. We can model these doublon dynamics with the following Hamiltonian:
\begin{equation}
\label{eq:H}
H=-J^0_{\text{Rb}}\sum_{ \langle \boldsymbol{i}, \boldsymbol{j} \rangle}  a_{\boldsymbol{i}}^{\dagger} a_{\boldsymbol{j}}- \sum_{\eta, \langle \boldsymbol{i}, \boldsymbol{j} \rangle} J_{\rm K}^\eta c_{\boldsymbol{i}, \eta}^{\dagger} c_{\boldsymbol{j},\eta}+ \sum_{\boldsymbol{i},\eta} U_\text{K-Rb}^\eta n_{\text{Rb},\boldsymbol{i}}^{0} n_{\text{K},\boldsymbol{i}}^\eta + U^0_{\text{Rb-Rb}}\sum_{\boldsymbol{i}} n_{\text{Rb},\boldsymbol{i}}^0 (n_{\text{Rb},\boldsymbol{i}}^0 -1),
\end{equation}
where $\eta=0$ and 1 denote respectively the ground and the first excited lattice bands. The first and second terms are the kinetic energy of the K and Rb atoms, respectively. Here, $a_{\boldsymbol{i}}(a_{\boldsymbol{i}}^\dagger)$ is the bosonic annihilation (creation) operator for a Rb atom at lattice site $\boldsymbol{i}$ in the lowest band, and $c_{\boldsymbol{i},\eta}(c_{\boldsymbol{i},\eta}^\dagger)$ is the fermionic annihilation (creation) operator for a K atom at lattice site $\boldsymbol{i}$ and band $\eta$.  We use $\langle \boldsymbol{i}, \boldsymbol{j} \rangle$ to indicate nearest-neighbor hopping between sites $\boldsymbol{i}$ and $\boldsymbol{j}$ with matrix element $J_{\alpha}^\eta$ with $\alpha=$ K or Rb. The third term describes the inter-species on-site interactions with matrix element $U_{\rm K-Rb}^\eta$. The last term is the on-site intra-species interaction between ground-band Rb atoms with strength $U^0_{\rm Rb-Rb}$, with $n_{\text{Rb},\boldsymbol{i}}^0$ as the occupation of site $\boldsymbol{i}$.
The tunneling rates and interaction energies are calculated for a particular $V_{\text{latt}}$ and $a_{\text{K-Rb}}$~\cite{Greiner2002}. For example, for $V_{\text{latt}}=10E_{\text{R}}$, $J_\text{K}^0/h=386$~Hz, $J_\text{Rb}^0/h=38.9$~Hz. The solid curves in Fig.~\ref{Figure3}a,b show the calculations based on the Hamiltonian given in Eq.~\eqref{eq:H}, where we have neglected Rb tunneling by setting $J_{\rm Rb}^0=0$.
We start with a single doublon, evolve the K for a hold time $\tau$, and then extract the doublon fraction from the probability that the K atom remains on the same site as the Rb atom.
In this treatment, we ignore the role of the magnetic-field sweeps. Calculations for a single doublon (solid lines), where the initial decay scales at $1-12(J_\text{K}^0/U^0_\text{K-Rb})^2$, agree well with the data, except at doublon fractions below $\sim30\%$ where the disagreement arises from the finite probability in the experiment that a K atom finds a different Rb partner. Simulating a Gaussian distribution of doublons with $10\%$ peak filling accounts for this effect (dashed lines) (see Methods).
The good agreement of these calculations with the data shows that tunneling of K, which is suppressed for deeper lattices, is the dominant mechanism for the reduction of the doublon fraction at short ($\sim$1~ms) times.  The on-site interaction with Rb suppresses the K tunneling when the interaction energy becomes larger than the width of the K Bloch band~\cite{Heinze2011}.

When studying doublon dynamics measured for two different initial atom conditions, we find indirect evidence for excited-band molecules.  
Here, we compare results for our usual molecule preparation using atomic insulators to a case where we start with a hotter initial atom gas mixture at a temperature above that for the Rb Bose-Einstein condensation (BEC) transition.  Using a band-mapping technique, we measure the initial population of K in the ground and first excited band, as shown in Figs.~\ref{Figure3}d $\text{\textbf{i}}$ and $\text{\textbf{ii}}$ (see Methods). We find that 11(2)$\%$ of the K atoms occupy the first excited band for the colder initial atom gas (these conditions are similar to those in Ref.~\cite{Moses2015} and are used in all the measurements described in this work, except for the green squares in Fig.~\ref{Figure3}c). When starting with the hotter atom gas, we measure a significantly higher K excited-band population of 31(6)$\%$.
When looking at doublon dynamics for these two cases (Fig.~\ref{Figure3}c), we observe a lower doublon fraction for the hotter initial gas for $V_{\text{latt}}\leq25E_{\text{R}}$. These data are taken for 1.68~mT/ms sweeps, $\tau=1$~ms, and $a_{\text{K-Rb}}=-220a_0$.

The lower doublon fraction can be explained by excited-band K atoms, which have a high tunneling rate ($J_\text{K}^0/h$ and $J_\text{K}^1/h$ are 89.3~Hz and 1110~Hz, respectively, for $V_{\text{latt}}=25E_{\text{R}}$). The presence of excited-band K atoms suggests that the $B$ sweeps for magneto-association (and dissociation) couple excited-band K atoms (plus a ground-band Rb atom) to excited-band Feshbach molecules. Moreover, the data suggest that the conversion efficiency for the excited-band Feshbach molecules is still high for $V_{\text{latt}}\leq25E_{\text{R}}$ since the measured increased fraction of the initial excited-band K atoms is similar to the observed difference (roughly $20\%$) in the doublon fraction (Fig.~\ref{Figure3}c). Since, in our preparation scheme, the doublons are directly formed from the dissociation of  ro-vibrational ground-state molecules, these results further indicate that a polar molecule sample prepared from a finite-temperature atom gas can contain a small fraction of molecules in an excited motional state in the lattice.  We note that we also observe a Rb excited-band population of 31(5)$\%$ after loading the thermal gas in the lattice; however, even for the excited band, the off-resonant Rb tunneling 
is slow compared to the 1~ms time scale of the measurements presented in Fig.~\ref{Figure3}.

The green dashed curve in Fig.~\ref{Figure3}c shows the model results for a K excited-band fraction of $24\%$. For comparison, the red solid curve, which is the same as the red curve in Fig.~\ref{Figure3}a,
includes no excited-band population.   For the hotter initial atom gas, the green dashed curve overlaps the data at the shallower lattice depths, but deviates from the measured doublon fraction at larger lattice depths. 
This may be expected since in the limit of a very deep lattice and a fully adiabatic magneto-association sweep, one expects that only the heavier atom (Rb) in excited-bands (plus a ground-band K atom) will couple to excited-band Feshbach molecules. Future studies 
of the magneto-association process in a lattice for systems such as K-Rb where center-of-mass and relative motion are coupled~\cite{Jachymski2013b} would be interesting and relevant to polar molecule preparation.

\begin{figure}
\centering
\includegraphics[width=7in]{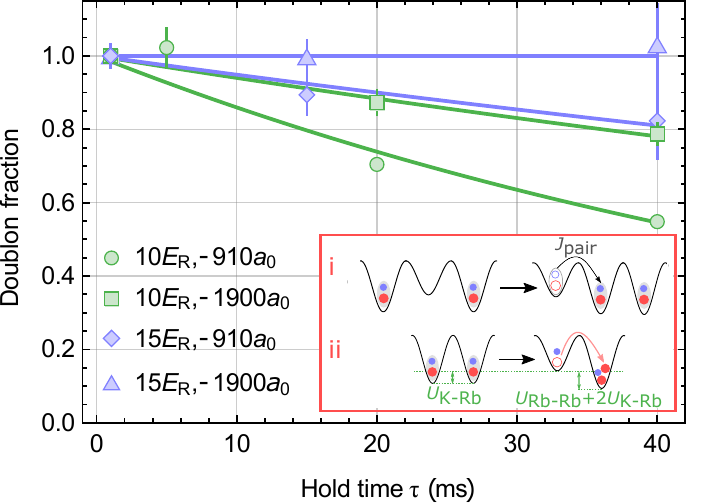}
\caption{\textbf{Longer time dynamics}. The dependence of the doublon fraction on the hold time $\tau$ in the lattice for both 10$E_\text{R}$ (green circles and squares) and 15$E_\text{R}$ (blue diamonds and triangles) for either $a_\text{K-Rb}=-910a_0$ (circles, diamonds) or $a_\text{K-Rb}=-1900a_0$ (squares, triangles). The lines are fits to exponential decay and are intended only as guides to the eye. (Inset) The doublon decay can involve tunneling of doublons through empty sites (\textbf{i}) prior to loss by the Rb tunneling process illustrated in (\textbf{ii}).}
\label{Figure4}
\end{figure}

In Fig.~\ref{Figure4}, we present data taken for $\tau$ up to 40 ms, in order to look for effects of Rb tunneling. Measurements of the remaining doublon fraction are shown for two lattice depths ($10E_{\text{R}}$ and $15E_{\text{R}}$) and two values of  $a_{\rm K-Rb}$ ($-910a_0$ and $-1900a_0$). In Fig.~\ref{Figure4}, the doublon fraction has been normalized by the measured value for $\tau=1$~ms in order to remove the effect of the shorter-time dynamics that were presented in Fig.~\ref{Figure3}a,b.  Similar to the shorter-time dynamics, at the longer hold times we observe a reduction in the doublon fraction that is suppressed for a deeper lattice and for strong inter-species interactions. Modeling these dynamics is theoretically challenging, and the lines in Fig.~\ref{Figure4} are exponential fits that are intended only as guides to the eye. Compared to doublons composed of identical bosons~\cite{Winkler2006} or fermions in two-spin states~\cite{Strohmaier2010}, the heteronuclear system has the additional complexities of two particle masses, two tunneling rates, and two relevant interaction energies. For example, for large $a_{\rm K-Rb}$, the interspecies interactions will strongly suppress Rb tunneling from a doublon to a neighboring empty site. Similarly, tunneling of a doublon to an empty lattice site is a slow second-order process at a rate $J_\text{pair}= 2J_\text{Rb}^0 J_\text{K}^0 /U_\text{K-Rb}^0$ due to the energy gap of $U_\text{K-Rb}^0$ (Fig.~\ref{Figure4} inset \textbf{i}).  However, Rb tunneling between two neighboring doublons, which creates a triplon (Rb-Rb-K) on one site and a lone K atom on the other site (Fig.~\ref{Figure4} inset \textbf{ii}), may occur on a faster time scale due to a much smaller energy gap of $U_\text{Rb-Rb}^0$ and the existence of the K tunneling band.  Considering these processes together, the time scale of the doublon decay would thus correspond to $J_\text{pair}$.
For $V_\text{latt}=10E_\text{R}$ and $a_\text{K-Rb}=-910a_0$, $J_\text{pair}=8$~Hz, which roughly matches the time scale of the decay observed in Fig.~\ref{Figure4}. These processes lead to many-body dynamics that depend on the filling fraction of doublons. For the purpose of creating polar molecules, we note that the observed time scale for these further dynamics is sufficiently long that they can be avoided in making molecules.

The studies discussed thus far demonstrate that  the Feshbach molecule conversion that we use to detect doublons could potentially underestimate the doublon fraction.
For example, the conversion efficiency of doublons containing excited-band atoms is complicated to calculate and is likely to be less than one. In addition, the efficiency of converting doublons to Feshbach molecules depends on the magnetic-field sweep rate, and as shown in Fig.~\ref{Figure2}c, a very slow sweep does not always yield a unity conversion efficiency. Finally, Feshbach molecules can suffer losses from inelastic collisions with other Feshbach molecules or unpaired K atoms~\cite{Ospelkaus2012}, which could reduce the measured number.
Given these factors and the importance of measuring the doublon fraction as a powerful diagnostic for optimizing molecule production from ultracold atoms in a lattice, we have implemented a second, complementary approach for measuring the doublon fraction using inelastic collisional loss instead of magneto-association.  
In our system, inelastic collisions are initiated by transferring the Rb atoms from the $\vert F=1, m_F=1 \rangle$ hyperfine to the $\vert 2, 2\rangle$ state, where $F$ is the total atomic spin and $m_F$ is its projection. Collisions of the $\vert 2, 2\rangle$ Rb atoms with K can result in spin relaxation back to the Rb $F = 1$ manifold. At $B=55$~mT, the $\vert 2, 2\rangle$ state is higher in energy by $h\times 8.1$~GHz, and this inelastic collision releases a large amount of energy compared to the trap depth and therefore results in atom loss from the trap. At a collision energy corresponding to 1 $\mu$K, the calculated inelastic collision rate using the coupled channels model of Ref.~\cite{Julienne2009} is $\beta = 6\times10^{-12}$~cm$^3$ s$^{-1}$, and using the on-site densities in a $V_{\text{latt}}=25E_{\text{R}}$ lattice, the resulting doublon lifetime is $\sim$2 ms.

\begin{figure}
\centering
\includegraphics[width=7in]{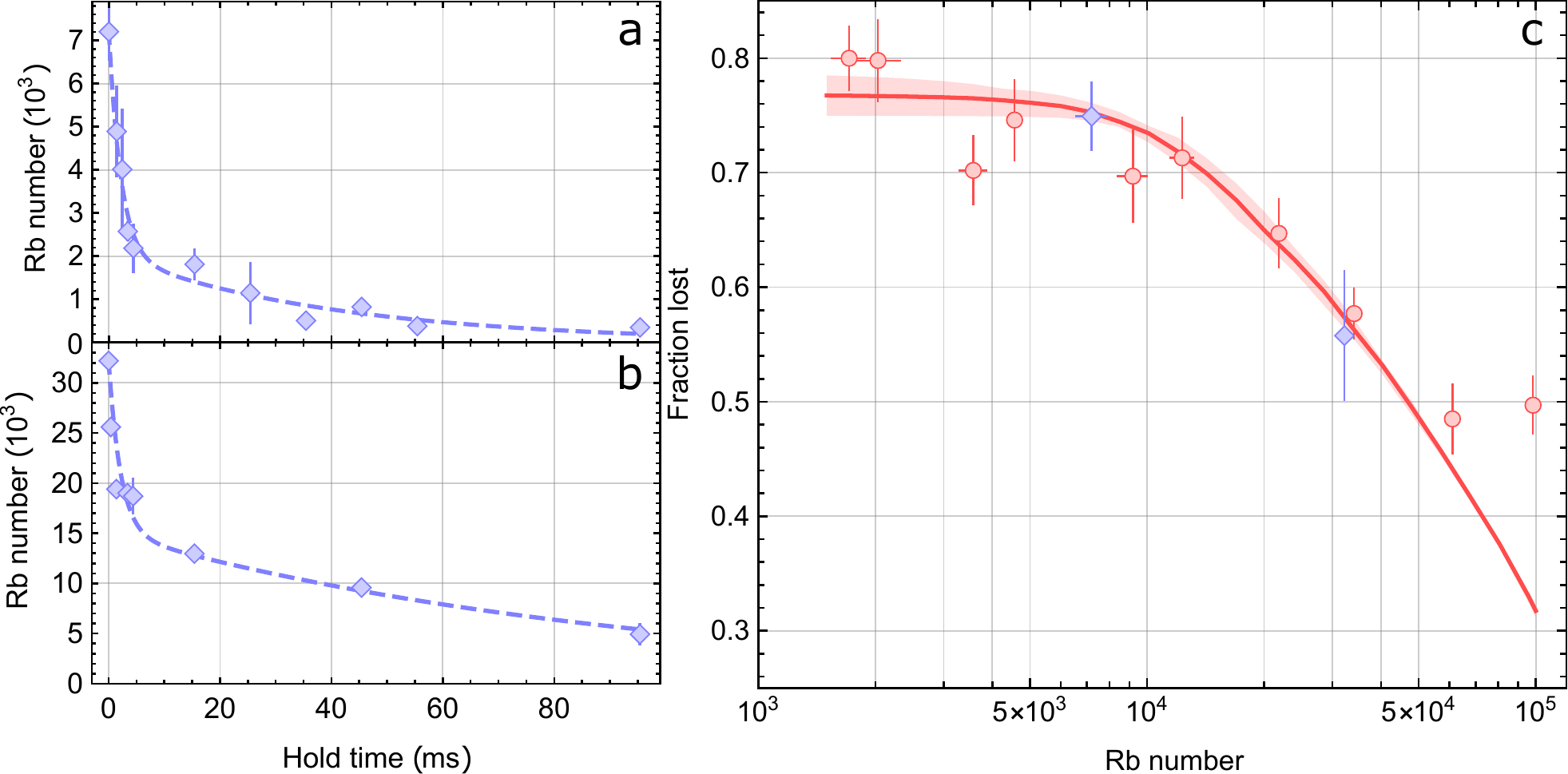}
\caption{\textbf{Measuring the initial atomic distributions with spin-changing collisions}. \textbf{a} and \textbf{b}, Sample data for low and high Rb number, respectively. The fraction of Rb remaining after the fast loss is different between the two cases. \textbf{c}, The fraction of Rb lost after $\sim8$~ms is plotted as a function of the Rb number in a 25$E_\text{R}$ lattice. At low Rb number, where the Mott insulator has one Rb atom per site, the fraction lost should be equal to the fraction of sites that have a K atom. As the Rb filling increases and the second Mott shell becomes populated, the fraction lost decreases. This technique yields both the filling of K and a measure of Rb atom number that corresponds to the onset of double occupancy of the Rb Mott insulator.}
\label{Figure5}
\end{figure}

In Fig.~\ref{Figure5}a and~\ref{Figure5}b, we show example data for the number of Rb atoms as a function of time after a 2.1 ms rf sweep that transfers Rb atoms to the $\vert 2, 2\rangle$ state. We observe a fast loss on the time scale of a few ms, followed by slower loss. We attribute the fast loss to inelastic collisions of Rb atoms in lattice sites shared with K, and the slow loss to tunneling of atoms followed by inelastic collisions. The dashed lines in Fig.~\ref{Figure5}a and~\ref{Figure5}b show a fit to the sum of two exponential decays with different time constants. We can extract the fraction of Rb that is lost on the short time scale from the fits.  We have compared this technique with Feshbach molecule formation, and we find that the two measurements generally agree.

As a further demonstration of the inelastic collision technique, we use this to probe the initial atomic distribution in the lattice before molecule formation.
Figure~\ref{Figure5}c shows the fraction of Rb atoms that are lost quickly from a $V_{\text{latt}}=25E_{\text{R}}$ lattice after the Rb atoms are transferred to the $\vert 2, 2\rangle$ state. For this data, we vary the initial number of Rb atoms that form a Mott insulator in the optical lattice prior to molecule creation.  In Fig.~\ref{Figure5}c, the blue diamonds correspond to the data shown in Fig.~\ref{Figure5}a and~\ref{Figure5}b. For the data shown in circles, the fraction lost is determined by comparing the Rb number measured before to that measured $8$ ms after the rf transfer.   
The solid curve shows a calculation of the expected loss for a Mott insulator with a temperature $T/J_\text{Rb}^0=15$ and a total radial harmonic confinement of 33~Hz for Rb. 
At low Rb number, where we expect only one Rb atom per site in the Mott insulator, the fraction lost is just the K filling fraction, assuming no double occupancy for K. For higher Rb number, double (and eventually triple and higher) occupancy in Mott shells causes a reduction in the fractional loss under the assumption of one Rb and one K lost per inelastic collision.  The shaded area indicates a $10\%$ uncertainty in the harmonic trapping frequency and $30\%$ uncertainty in $T$. From the fit, we extract a K filling fraction of $0.77(2)$, which is in excellent agreement with the measured peak K filling reported in Ref.~\cite{Moses2015}. We note the previously measured fraction of Rb converted to Feshbach molecules at low Rb atom number was significantly less than $0.80$~\cite{Moses2015}. This disagreement may now be attributed to the large number of K atoms present in the lattice after molecule formation, which can induce losses through inelastic collisions, and to the effect of 
the $d$-wave resonance when making molecules as discussed above.

Our investigation of heteronuclear doublons and their conversion to molecules by magneto-association reveals 
the important roles played by the lattice depth for both atomic species, the inter-species interactions, the population in excited motional states of the lattice, and the magnetic-field sweep rate. The study of doublon dynamics provides insight into the optimal preparation conditions for polar molecules in a 3D optical lattice. The highly non-equilibrium state of doublons that we use for these studies also provides an intriguing system for exploring Hubbard dynamics of a Bose-Fermi mixture, where the behavior of the many-body system can depend on two different tunneling rates and two different interaction strengths~\cite{Best2009,Heinze2011}. 
Further investigations of this system could include going to 1D where experiments can be benchmarked with theory and pursuing quantum simulation of novel phases such as quasi-crystalization and many-body localization in higher dimensions~\cite{Gopalakrishnan2013, Martin2015,Schreiber2015,Monroe2015}.

\bibliographystyle{unsrt}

\begin{thebibliography}{10}

\bibitem{Baranov2008}
M.A. Baranov.
\newblock Theoretical progress in many-body physics with ultracold dipolar
  gases.
\newblock {\em Physics Reports}, 464(3):71 -- 111, 2008.

\bibitem{Pupillo2009}
G.~Pupillo, A.~Micheli, H.~P. B\"uchler, and P.~Zoller.
\newblock Condensed matter physics with cold polar molecules.
\newblock {\em Book Chapter in 'Cold Molecules: Theory, Experiment, and
  Applications'}, CRC Press, 2009.

\bibitem{Carr2009}
Lincoln~D. Carr, David DeMille, Roman~V. Krems, and Jun Ye.
\newblock {Cold and ultracold molecules: Science, technology and applications}.
\newblock {\em New Journal of Physics}, 11:0--87, 2009.

\bibitem{Lahaye2009}
T~Lahaye, C~Menotti, L~Santos, M~Lewenstein, and T~Pfau.
\newblock {The physics of dipolar bosonic quantum gases}.
\newblock {\em Review of Progress in Physics}, 126401:71, 2009.

\bibitem{Lemeshko2013}
Mikhail Lemeshko, Roman~V. Krems, John~M. Doyle, and Sabre Kais.
\newblock Manipulation of molecules with electromagnetic fields.
\newblock {\em Molecular Physics}, 111:1648--1682, 2013.

\bibitem{Yao2012}
N.~Y. Yao, C.~R. Laumann, A.~V. Gorshkov, S.~D. Bennett, E.~Demler, P.~Zoller,
  and M.~D. Lukin.
\newblock Topological flat bands from dipolar spin systems.
\newblock {\em Phys. Rev. Lett.}, 109:266804, Dec 2012.

\bibitem{Gorshkov2011}
Alexey~V. Gorshkov, Salvatore~R. Manmana, Gang Chen, Jun Ye, Eugene Demler,
  Mikhail~D. Lukin, and Ana~Maria Rey.
\newblock Tunable superfluidity and quantum magnetism with ultracold polar
  molecules.
\newblock {\em Phys. Rev. Lett.}, 107:115301, Sep 2011.

\bibitem{Syzranov2014}
Sergey~V Syzranov, Michael~L Wall, Victor Gurarie, and Ana~Maria Rey.
\newblock {Spin-orbital dynamics in a system of polar molecules}.
\newblock {\em Nature Communications}, 5:10, 2014.

\bibitem{Yan2013}
Bo~Yan, Steven~A. Moses, Bryce Gadway, Jacob~P. Covey, Kaden R.~A. Hazzard,
  Ana~Maria Rey, Deborah~S. Jin, and Jun Ye.
\newblock Observation of dipolar spin-exchange interactions with
  lattice-confined polar molecules.
\newblock {\em Nature}, 501(7468):521--525, 09 2013.

\bibitem{Hazzard2014}
Kaden R.~A. Hazzard, Bryce Gadway, Michael Foss-Feig, Bo~Yan, Steven~A. Moses,
  Jacob~P. Covey, Norman~Y. Yao, Mikhail~D. Lukin, Jun Ye, Deborah~S. Jin, and
  Ana~Maria Rey.
\newblock Many-body dynamics of dipolar molecules in an optical lattice.
\newblock {\em Phys. Rev. Lett.}, 113:195302, Nov 2014.

\bibitem{Moses2015}
Steven~A. Moses, Jacob~P. Covey, Matthew~T. Miecnikowski, Bo~Yan, Bryce Gadway,
  Jun Ye, and Deborah~S. Jin.
\newblock Creation of a low-entropy quantum gas of polar molecules in an
  optical lattice.
\newblock {\em arXiv:1507.02377}, 2015.

\bibitem{Safavi-Naini2015}
A.~Safavi-Naini, M.~L. Wall, and Ana~Maria Rey.
\newblock {The Role of Interspecies Interactions in the Preparation of a
  Low-entropy Gas of Polar Molecules in a Lattice}.
\newblock {\em ArXiv:1507.03697}, 2015.

\bibitem{Winkler2006}
K.~Winkler, G.~Thalhammer, F.~Lang, R.~Grimm, J.~Hecker Denschlag, A.~J. Daley,
  A.~Kantian, H.~P. B\"uchler, and P.~Zoller.
\newblock Repulsively bound atom pairs in an optical lattice.
\newblock {\em Nature}, 441(7311):853--856, 06 2006.

\bibitem{Zaccanti2006}
M.~Zaccanti, C.~D'Errico, F.~Ferlaino, G.~Roati, M.~Inguscio, and G.~Modugno.
\newblock Control of the interaction in a fermi-bose mixture.
\newblock {\em Phys. Rev. A}, 74:041605, Oct 2006.

\bibitem{Julienne2009}
P~S Julienne.
\newblock {Ultracold molecules from ultracold atoms: a case study with the KRb
  molecule}.
\newblock {\em Faraday discussions}, 142:361--388, 2009.

\bibitem{Bloom2013}
Ruth~S. Bloom, Ming~Guang Hu, Tyler~D. Cumby, and Deborah~S. Jin.
\newblock {Tests of universal three-body physics in an ultracold Bose-Fermi
  Mixture}.
\newblock {\em Physical Review Letters}, 111(10):1--5, 2013.

\bibitem{Ruzic2013}
Brandon~P. Ruzic, Chris~H. Greene, and John~L. Bohn.
\newblock {Quantum defect theory for high-partial-wave cold collisions}.
\newblock {\em Physical Review A - Atomic, Molecular, and Optical Physics},
  87(3):1--14, 2013.

\bibitem{Thalhammer2006}
G.~Thalhammer, K.~Winkler, F.~Lang, S.~Schmid, R.~Grimm, and J.~Hecker
  Denschlag.
\newblock Long-lived feshbach molecules in a three-dimensional optical lattice.
\newblock {\em Phys. Rev. Lett.}, 96:050402, Feb 2006.

\bibitem{Greif2011}
Daniel Greif, Leticia Tarruell, Thomas Uehlinger, Robert J\"{o}rdens, and
  Tilman Esslinger.
\newblock {Probing nearest-neighbor correlations of ultracold fermions in an
  optical lattice}.
\newblock {\em Physical Review Letters}, 106(14):1--4, 2011.

\bibitem{Chotia2012}
Amodsen Chotia, Brian Neyenhuis, Steven~A. Moses, Bo~Yan, Jacob~P. Covey,
  Michael Foss-Feig, Ana~Maria Rey, Deborah~S. Jin, and Jun Ye.
\newblock Long-lived dipolar molecules and feshbach molecules in a 3d optical
  lattice.
\newblock {\em Phys. Rev. Lett.}, 108:080405, Feb 2012.

\bibitem{Mies2000}
F.~H. Mies, E.~Tiesinga, and P.~S. Julienne.
\newblock Manipulation of feshbach resonances in ultracold atomic collisions
  using time-dependent magnetic fields.
\newblock {\em Phys. Rev. A}, 61:022721, 2000.

\bibitem{Julienne2004withnote}
Paul~S. Julienne, Eite Tiesinga, and Thorsten Köhler.
\newblock Making cold molecules by time-dependent feshbach resonances.
\newblock {\em Journal of Modern Optics}, 51(12):1787--1806, 2004.
\newblock We note the right-hand side of Equation (26) in this reference is
  missing a factor of $\pi$. This error was corrected in Ref. arXiv:0312492v3.

\bibitem{Klempt2008}
C.~Klempt, T.~Henninger, O.~Topic, M.~Scherer, L.~Kattner, E.~Tiemann,
  W.~Ertmer, and J.~J. Arlt.
\newblock Radio-frequency association of heteronuclear feshbach molecules.
\newblock {\em Phys. Rev. A}, 78:061602, 2008.

\bibitem{Jurgensen2014}
Ole J\"{u}rgensen, Florian Meinert, Manfred~J. Mark, Hanns-Christoph
  N\"{a}gerl, and Dirk-S\"{o}ren L\"{u}hmann.
\newblock {Observation of Density-Induced Tunneling}.
\newblock {\em Physical Review Letters}, 113(19):1--5, 2014.

\bibitem{Greiner2002}
Markus Greiner, Olaf Mandel, Tilman Esslinger, Theodor~W. H\"{a}nsch, and
  Immanuel Bloch.
\newblock {Quantum phase transition from a superfluid to a Mott insulator in a
  gas of ultracold atoms.}
\newblock {\em Nature}, 415(6867):39--44, 2002.

\bibitem{Heinze2011}
J.~Heinze, S.~G\"{o}tze, J.~S. Krauser, B.~Hundt, N.~Fl\"{a}schner, D.~S.
  L\"{u}hmann, C.~Becker, and K.~Sengstock.
\newblock {Multiband spectroscopy of ultracold fermions: Observation of reduced
  tunneling in attractive Bose-Fermi mixtures}.
\newblock {\em Physical Review Letters}, 107(13):21--25, 2011.

\bibitem{Jachymski2013b}
Krzysztof Jachymski, Zbigniew Idziaszek, and Tommaso Calarco.
\newblock Feshbach resonances in a nonseparable trap.
\newblock {\em Phys. Rev. A}, 87:042701, 2013.

\bibitem{Strohmaier2010}
Niels Strohmaier, Daniel Greif, Robert J\"{o}rdens, Leticia Tarruell, Henning
  Moritz, Tilman Esslinger, Rajdeep Sensarma, David Pekker, Ehud Altman, and
  Eugene Demler.
\newblock {Observation of elastic doublon decay in the fermi-hubbard model}.
\newblock {\em Physical Review Letters}, 104(8):1--4, 2010.

\bibitem{Ospelkaus2012}
S.~Ospelkaus, K.-K. Ni, D.~Wang, M.~H.~G. de~Miranda, B.~Neyenhuis,
  G.~Qu\'{e}m\'{e}ner, P.~S. Julienne, J.~L. Bohn, D.~S. Jin, and J.~Ye.
\newblock Quantum-state controlled chemical reactions of ultracold
  potassium-rubidium molecules.
\newblock {\em Science}, 327(5967):853--857, 2010.

\bibitem{Best2009}
Th~Best, S.~Will, U.~Schneider, L.~Hackerm\"{u}ller, D.~{Van Oosten}, I.~Bloch,
  and D.~S. L\"{u}hmann.
\newblock {Role of interactions in Rb87-K40 Bose-Fermi mixtures in a 3D optical
  lattice}.
\newblock {\em Physical Review Letters}, 102(3):100--103, 2009.

\bibitem{Gopalakrishnan2013}
Sarang Gopalakrishnan, Ivar Martin, and Eugene~A. Demler.
\newblock Quantum quasicrystals of spin-orbit-coupled dipolar bosons.
\newblock {\em Phys. Rev. Lett.}, 111:185304, 2013.

\bibitem{Martin2015}
Ivar Martin, Sarang Gopalakrishnan, and Eugene~A. Demler.
\newblock {Weak crystallization theory of metallic alloys}.
\newblock {\em arXiv:1506.03077v1}, 2015.

\bibitem{Schreiber2015}
Michael Schreiber, Sean~S. Hodgman, Pranjal Bordia, Henrik~P. Lüschen, Mark~H.
  Fischer, Ronen Vosk, Ehud Altman, Ulrich Schneider, and Immanuel Bloch.
\newblock Observation of many-body localization of interacting fermions in a
  quasirandom optical lattice.
\newblock {\em Science}, 349(6250):842--845, 2015.

\bibitem{Monroe2015}
Jacob Smith, Aaron Lee, Philip Richerme, Brian Neyenhuis, Paul~W. Hess, Philipp
  Hauke, Markus Heyl, David~A. Huse, and Christopher Monroe.
\newblock {Many-body localization in a quantum simulator with programmable
  random disorder}.
\newblock {\em ArXiv:1508.07026}, 2015.

\bibitem{Jachymski2013a}
Krzysztof Jachymski and Paul~S. Julienne.
\newblock Analytical model of overlapping feshbach resonances.
\newblock {\em Phys. Rev. A}, 88:052701, 2013.

\bibitem{Murthy2014}
P.~A. Murthy, D.~Kedar, T.~Lompe, M.~Neidig, M.~G. Ries, A.~N. Wenz, G.~Z\"urn,
  and S.~Jochim.
\newblock Matter-wave fourier optics with a strongly interacting
  two-dimensional fermi gas.
\newblock {\em Phys. Rev. A}, 90:043611, 2014.

\end{thebibliography}

\textbf{Acknowledgments} We thank M. Wall for useful discussions. We acknowledge funding for this work from NIST, NSF grant number 1125844, AFOSR-MURI and ARO-MURI. J.P.C. acknowledges funding from the NDSEG fellowship. Part of the computation for this work was performed at the University of Oklahoma Supercomputing Center for Education and Research (OSCER).

\textbf{Contributions} The experimental work and data analysis were carried out by J.P.C., S.A.M., M.T.M., Z.F., D.S.J. and J.Y. Theoretical modeling and calculations are provided by M.G., A.S.-N., J.S., P.S.J., and A.M.R. All authors discussed the results and contributed to the preparation of the manuscript.

\textbf{Competing Interests} The authors declare that they have no competing financial interests.

\textbf{Correspondence} Correspondence should be addressed to Deborah Jin~(email: jin@jilau1.colorado.edu) and Jun Ye~(email: ye@jila.colorado.edu).

\vspace{1cm}
\textbf{Methods}

\textbf{Optical trapping potentials}  The preparation of the atomic gas in a 3D lattice, with a wavelength of $1064$~nm, as well as the creation of ground-state polar molecules, follows the procedures described in Ref.~\cite{Moses2015}. The lattice is superimposed on a crossed-beam optical dipole trap that is cylindrically symmetric. The dipole trap alone has an axial trap frequency of $\omega_z = 2\pi \times 180$ Hz in the vertical direction and a radial trap frequency of $\omega_r = 2\pi \times 25$ Hz for Rb. The measured optical trap frequencies for K are $2\pi \times 260$ Hz and $2\pi \times 30$ Hz.

\textbf{Width of the $d$-wave resonance}
The scattering lengths reported in Fig.~\ref{Figure3} have been calculated using $a_{\text{K-Rb}}(B)=a_{bg}[1-\Delta_s/(B-B^{\text{res}}_s)]$ with $a_{bg}=-187(5)a_0$, $B^{\text{res}}_s=54.662\,$mT, and $\Delta_s=0.304\,$mT~\cite{Klempt2008}.
Including the $d$-wave resonance, the scattering length can be parametrized by $a_{\text{K-Rb}}(B)=a_{bg}[1-\Delta_s/(B-B^{\text{res}}_s)-\Delta_d/(B-B^{\text{res}}_d)]$ ~\cite{Jachymski2013a}. Exploiting that $\Delta_d \ll \Delta_s$, we can write $a_{\text{K-Rb}}(B)\approx a_{bg}^\prime[1-\Delta_d^\prime/(B-B^{\text{res}}_d)]$ near the $d$-wave resonance, where $a_{bg}^\prime=a_{bg}[1-\Delta_s/(B^{\text{res}}_d-B^{\text{res}}_s)]$ and $\Delta_d^\prime=\Delta_d/[1-\Delta_s/(B^{\text{res}}_d-B^{\text{res}}_s)]$. This has the form of an isolated resonance and we can apply the findings of Ref.~\cite{Julienne2004withnote}, namely that $A=4\sqrt{3}\omega_{\text{HO}}|a_{bg}^\prime\Delta_d^\prime|/L_{\text{HO}}$, to determine $\Delta_d=a_{bg}^\prime \Delta_d^\prime/a_{bg}=A L_{\text{HO}}/(4\sqrt{3}\omega_{\text{HO}}a_{bg}) $. We note that Ref.~\cite{Ruzic2013} predicts $\Delta_d'=6.3\times10^{-4}$~mT, which is larger than our determination of $\Delta_d'=2.0(2)\times10^{-4}$~mT.

In this determination, we ignore the coupling between center of mass and relative motion that arises from the fact that K and Rb experience different trapping potentials in the optical lattice. We use an effective trap frequency $\omega_{\text{HO}}=\sqrt{\left( m_{\text{Rb}}\omega_{\text{K}}^2+m_{\text{K}}\omega_{\text{Rb}}^2\right)/\left(m_{\text{Rb}}+m_{\text{K}} \right)}$ that governs the dynamics in the relative coordinate. Here, $m_{\text{Rb}}$ and $m_{\text{K}}$ are the masses of the Rb and K atom, respectively. The trap frequency for Rb is given by $\omega_{\text{Rb}}=2(E_{\text{R}}/\hbar)\sqrt{V_{\text{latt}}/E_{\text{R}}}$, and for the 1064~nm lattice, the trap frequency for K is $\omega_{\text{K}}\approx1.4\omega_{\text{Rb}}$. For $V_\text{latt}=35E_\text{R}$, $\omega_\text{HO}=30.4$~kHz.

\textbf{Density distribution of doublons}
The dashed lines in Fig.~\ref{Figure3}a,b have been obtained by random sampling of initial doublon positions according to a Gaussian probability distribution of the filling fraction. A peak filling of 10$\%$ and widths $\sigma_x=\sigma_y=6.5\sigma_z=21$ sites have been used, corresponding to $N\approx 2000$ doublons. The experimentally determined cloud sizes are slightly larger ($\sigma_x=25$ to $42$ sites) but we checked that the resulting doublon fraction is converged with respect to the cloud size. In-situ absorption images of the cloud are consistent with a Gaussian distribution of 5-10$\%$ peak filling. Tunneling of Rb is neglected in the model, where initially each K atom is localized on site containing a Rb atom and the doublon fraction is defined as the probability to find the K atom on a site with Rb after the evolution time $\tau$.

\textbf{Band mapping} To measure the excited-band fraction of the initial K atoms, we use a band-mapping technique (Fig.~\ref{Figure3}d). Starting with the K atoms in the 3D lattice plus optical dipole trap potential, we turn off the lattice in 1~ms and allow the the K gas to expand in the optical dipole trap for a quarter trap period~\cite{Murthy2014}. We image the cloud with a probe beam that propagates along the vertical direction.

\end{document}